\documentclass[12pt]{article}
\def\appendix{\par
  \setcounter{section}{0}%
  \def\@chapapp{APPENDIX}%
  \def\thesection{Appendix \Alph{section}:}
  }

\usepackage{amsmath,amssymb,graphicx,slashed}
\usepackage[all]{xy}

\newcommand{\eweak}{$\mathrm{SU(2)\otimes U(1)}\ $}
\newcommand{\custodial}{$\mathrm{SU(2)_L\otimes SU(2)_R}\ $}
\newcommand{\zbb}{$Z\rightarrow\overline{b}_L b_L\ $}
\newcommand{\half}{\frac{1}{2}}

\newcommand{\doublerightxyarrow}{\ar@{-}[rr] |-{\SelectTips{eu}{}\object@{>}}}
\newcommand{\rightxyarrow}{\ar@{-}[r] |-{\SelectTips{eu}{}\object@{>}}}
\newcommand{\doublerightDDxyarrow}{\ar@/_1pc/@{-}[rr] |-{\SelectTips{eu}{}\object@{>}}}
\newcommand{\doublerightDxyarrow}{\ar@/_0.3pc/@{-}[rr] |-{\SelectTips{eu}{}\object@{>}}}
\newcommand{\doublerightUxyarrow}{\ar@/^0.3pc/@{-}[rr] |-{\SelectTips{eu}{}\object@{>}}}
\newcommand{\doublerightUUxyarrow}{\ar@/^1pc/@{-}[rr] |-{\SelectTips{eu}{}\object@{>}}}

\begin{document}
\title{\bf The Advantages of Four Dimensions for Composite Higgs Models}
\author{Matthew Baumgart\footnote{baumgart@fas.harvard.edu}\\
\small\sl Jefferson Physical Laboratory \\
\small\sl Harvard University \\
\small\sl Cambridge, MA 02138}
\date{}
\maketitle

\begin{abstract}
We examine the relationship between little Higgs and 5d composite models with identical symmetry structures.  By performing an ``extreme" deconstruction, one can reduce any warped composite model to a little Higgs theory on a handful of sites.  This allows us to use 4d intuition and the powerful constraints of nonlinear sigma models to elucidate obscure points in the original setup.  We find that the finiteness of the Higgs potential in 5d is due to the same collective symmetry breaking as in the little Higgs.  We compare a 4d and a 5d model with the same symmetry to the data.  Reviewing the constraints on models related to the Minimal Composite Higgs (hep-ph/0412089), we see that it has difficulty in producing acceptable values for $S,\ T$, and $m_\mathrm{top}$ simultaneously.  By contrast, in a global analysis, the Minimal Moose with custodial symmetry is viable in a large region of its parameter space and suffers from no numeric tunings.  We conjecture that this result is generic for 4d and 5d models with identical symmetries.  The data will less strongly constrain the little theory.     
\end{abstract}

\section{Introduction}

\hspace{0.15in} The Standard Model is famously plagued by the need for a light scalar.  One possible solution is that the Higgs is a pseudo-Goldstone \cite{GK}.  This idea has received renewed attention with little Higgs theories \cite{Arkani-Hamed:2001nc, Arkani-Hamed:2002pa, Arkani-Hamed:2002qx, Arkani-Hamed:2002qy} and certain 5d warped models \cite{Contino:2003ve, Agashe:2004rs, Carena:2006bn, Contino:2006qr}.  In the former, the Higgs is protected by approximate, nonlinearly realized symmetries.  The protection is redundant, such that breaking any individual symmetry will not generate a potential for the Higgs.  This collective symmetry breaking prevents the generation of a quadratically  divergent Higgs mass.  In the 5d setup, the Higgs is the bulk $A^5$ zero mode whose holographic dual is a 4d PGB arising from strong CFT dynamics.  The protection of the Higgs potentials in these models is usually ascribed to the need for brane-to-brane propagators.

In fact, it is just collective symmetry breaking.  We can see this by deconstructing the fifth dimension \cite{Arkani-Hamed:2001ca} and integrating out the heavy modes.  This gives us an extreme deconstruction where the theory now lives on a small number of sites.  Counting global and gauged symmetries shows that we have produced a little Higgs theory.  As an example, we will study the Minimal Composite Higgs Model (MCHM) and its descendants \cite{Agashe:2004rs, Carena:2006bn, Contino:2006qr}.  This warped model has an SO(5) bulk gauge symmetry and SO(4) and \eweak gauge symmetries on the IR and UV branes, respectively.  By deconstructing this theory and running to a low scale, we recover the symmetry structure of the Custodial Minimal Moose (CMM) \cite{Chang:2003un}.  In this, there are two sites, one with SO(4) and one with \eweak gauge symmetry.

PGB Higgs models such as these represent a large and interesting class of beyond the Standard Model theories.  It becomes an important question to decide which realistic models look the most promising.  Comparing MCHM to the CMM will help to answer whether we do better to place a given symmetry pattern on branes in 5d or on sites in 4d.  Both models are suppressing contributions to $T$ and nonoblique corrections.  MCHM has a known moderate tuning for $S$ of $\mathcal{O}(10\%)$, but were this the only difficulty we may forgive it in exchange for a UV complete theory (to the Planck scale) with natural flavor hierarchy.  In fact, we will see that the $S$ parameter and changes to the \zbb rate force the theory into a corner of parameter space where reproducing $m_\mathrm{top}$ conflicts with getting acceptable precision electroweak observables.  

The CMM has no large contributions to $S$ and uses custodial SU(2) to suppress $T$ and the nonobliques.  Thus, its parameter space is free of tunings and tensions.  It can even have the scale of new physics as low as 375 GeV.  We also have the freedom to add eight new quarks without causing problems for $S$ or tuning in the Higgs potential.  Doing so allows us to generate weak scale dark matter within the CMM as discussed in Section \ref{sec:dmv}.  

The structure of the paper is the following:  In Section \ref{sec:appDecWarp}, we will show how to deconstruct a 5d warped model and what one can learn from doing so.  We will take an example from the Higgs as a Holographic PGB \cite{Contino:2003ve}.  In it, a possibly natural mechanism was proposed for generating an $\mathcal{O}(1)$ Higgs quartic with no mass term.  We show that the method used actually requires tuning.  We also present an argument that in 5d what protects the Higgs potential is just collective symmetry breaking.  Section \ref{sec:cmm} is a review of the Custodial Minimal Moose and a discussion of possible modifications to the third generation for protecting \zbb or generating dark matter.  In Section \ref{sec:PEWk} we review the constraints on MCHM and present a global analysis of the CMM.  Section \ref{sec:conc} is the Conclusion. 

\section{Applying Deconstruction to Warped Models}
\label{sec:appDecWarp}

\hspace{0.15in} The little Higgs began as a deconstruction of higher dimensional theories \cite{Arkani-Hamed:2001nc, Arkani-Hamed:2002pa}.  Since deconstruction rewrites a theory into four dimensions, we can use it to clarify points which may seem obscure in the higher dimensional framework.  Furthermore, it allows us to systematically convert warped setups into little Higgs models.  By doing so, we can eliminate the large $N$ CFT that forces tuning upon warped models by their contributions to the $S$ parameter.  Perturbativity is retained as our 4d field theory is now weakly coupled.  As we shall see in this section, the viability of MCHM as a PGB Higgs theory without a hierarchy problem can be seen purely from its symmetry pattern.  We can also decide how symmetry breakings in the UV will manifest themselves at low energies.

\subsection{Tuning the Holographic Higgs}

\hspace{0.15in} In this 5d model \cite{Contino:2003ve}, an $\mathrm{SU(3)_L}$ gauge symmetry lives in the bulk.  It is broken to $\mathrm{SU(2)_L}$ by two separate SU(3) triplet scalars, one on each brane.  As the UV brane scalar gets a Planck scale vev, the PGB Higgs is almost entirely the IR field.  This takes the form of excitations about the $\mathrm{SU(2)_L}$ preserving vacuum configuration $\left( \begin{array}{ccc} 0 & 0 & f \end{array} \right)$.  We can thus parameterize the Higgs field as follows:

\begin{equation}
\Sigma=e^{\frac{i}{f}\pi^a t^a}\left( \begin{array}{c} 0 \\
																											 0 \\
																											 f \end{array} \right) = f \left( \begin{array}{c} \frac{\sqrt{2}ih^+}{h}\sin\left(\frac{h}{\sqrt{2}f}\right) \\
																											 \frac{\sqrt{2}ih^0}{h}\sin\left(\frac{h}{\sqrt{2}f}\right) \\
																											 \cos\left(\frac{h}{\sqrt{2}f}\right) \end{array} \right),
\end{equation}

\vspace{0.15in} \noindent where the $\pi^a$ are the SU(3)/SU(2) generators.  There was a proposal to generate a tree-level quartic and vanishing mass term with a bulk scalar $\Phi$ transforming as a $\mathbf{6}$ of $\mathrm{SU(3)}$.\footnote{In \cite{Agashe:2004rs}, a similar method was used to add terms to the Higgs potential.  However, there the goal was more modest.  The authors wished only to deform the radiative Higgs potential to alleviate the tuning required by $S$.}  Using deconstruction, we can see that such a potential is necessarily tuned.  The Higgs couples to $\Phi$ with the interaction, $\Sigma^\top \Phi^\dagger \Sigma$.  Under $\mathrm{SU(2)}$, the $\Phi$ will decompose as triplet, doublet, and singlet. By integrating out $\Phi$, one induces a Higgs potential.   Assuming an appropriate symmetry breaking pattern on the UV brane, the doublet and singlet contributions to the Higgs mass can cancel, leaving just the quartic.  By deconstructing the theory and introducing spurions, we can check if, in fact, there is a way to engineer such a cancellation.  The collective symmetry breaking is exactly the same as in the Simplest Little Higgs \cite{Schmaltz:2004de}.  We can express the theory at some scale $q$ with $M_P \gg q \gg \mathrm{TeV}$ as the following moose (the procedure for turning warped models into mooses is explained in the next subsection):

\begin{equation}
\begin{tabular}{c}
\xymatrix@R=.4pc@C=1.4pc{\mathrm{Global:} & SU(3) && \mathrm{SU(3)^2} &&  \\
& *=<20pt>[o][F]{} \doublerightxyarrow && *=<20pt>[o][F]{} \doublerightxyarrow^{\mbox{\raisebox{1.5ex}{$\Sigma$}}} && *=<20pt>[o][]{} \\ \mathrm{Gauged:} & SU(2) && SU(3) && }
\end{tabular}
\end{equation}

\vspace{0.15in} \noindent By the rightmost ``link," we mean simply that $\Sigma$ transforms as a $\mathbf{3}$ of SU(3) under the symmetries of that site.  The use of $\Phi$ in the Holographic Higgs was to communicate a certain pattern of symmetry breaking on the UV brane to the Higgs on the IR brane.  In the deconstructed version, this communication will necessarily take the form: 
\begin{equation*}
\mathrm{\left(IR\ site\ fields\right)\times\left(link\right)\times\left(\mathrm{SU(3)}\ breaking\ operators\right)\times\left(link\right)\times\left(IR\ site\ fields\right)}
\end{equation*}
\noindent We can represent the above chain as an operator that transforms under the $\mathrm{SU(3)}$ symmetry of the IR site.  At low energies, it is reduced to being a spurion, and all interactions must remain invariant under the residual $\mathrm{SU(2)}$.  In fact, there is only one type of nontrivial operator involving $\Sigma$ we can write down, $\left(\Sigma^\dagger B \Sigma\right)^n$, where  

\begin{eqnarray}
B &=& \left( \begin{array}{ccc} a & 0 & 0 \\
															0 & a & 0 \\
															0 & 0 & b	
\end{array} \right),\ \mathrm{and}\\
B&\rightarrow& U B U^\dagger\;\mathrm{under\ SU(3)}.
\end{eqnarray}

\noindent Expanding out the $n=1$ case, we get 

\begin{equation}
V(h)_{n=1}=\lambda(a-b)h^2+\frac{\lambda(b-a)}{3f^2}h^4,
\end{equation}

\noindent where $\lambda$ is an overall coupling of mass dimension 1.  Thus, with this term it is not possible to cancel the mass term without affecting the quartic as well.  We could also include the $n=2$ contribution:

\begin{equation}
V(h)_{n=2}=\frac{1}{M^2}\left(2b\:(a-b)f^2 h^2+\frac{1}{3}(3a-5b)\:(a-b)h^4 \right),
\end{equation}

\noindent with $M$ some heavy mass on the IR site that induces this operator.  We can imagine tuning the two contributions against each other.  However, that will not be radiatively stable.  This is the exact problem one encounters in the Simplest Little Higgs \cite{Schmaltz:2005ky}.   

\subsection{Why There is No Hierarchy Problem in MCHM}

\hspace{0.15in} In the original MCHM papers, the absence of quadratic divergences in the Higgs radiative potential is justified in five dimensions.  In this version of the theory, the Higgs is the zero mode of the bulk gauge field, $A^5$.  Clearly, in the infinite bulk limit, it will have no self-interaction potential.  After adding boundaries, we can still have the zero mode by imposing Neumann boundary conditions on both branes.  Thus, contributions to the potential necessarily involve brane to brane propagators.  The claim is that since these are long-distance and nonlocal, they are finite.  However, this is bizarre from the point of view of holographic RG.  Integrating out high energy modes is equivalent to moving the UV brane toward the IR one, while inducing local operators on the former \cite{NAH:2007}.  Thus, as we integrate out modes down to the weak scale, we lose the ability to credit the finiteness to a long-distance effect.  The UV brane is now only a short distance from the IR.  

To perform the integration, we deconstuct the fifth dimension into of a large number of sites.  For simplicity, let us consider a simple example with a $\mathrm{U(1)}$ gauge theory in the bulk and $\mathrm{U(1)}$ on both branes.  The setup is the following.  We use a constant lattice spacing, $a$, and a uniform bulk gauge coupling, $g_5$.  The 4d gauge coupling, $g$, will also be constant across sites.  The link vevs, $f_i$, change exponentially with scale, providing the remnant of the warp factor.  Following \cite{Randall:2002qr}, we have

\begin{equation}
f_j=\frac{e^{-ky_j}}{a}.
\end{equation}

\noindent The gauge boson mass matrix for this setup will be:

\begin{equation}
M^2=	\left( \begin{array}{ccccc} 
\vspace{0.08in} g^2f_1^2 && -g^2f_1^2 && 0 \\
\vspace{0.02in} -g^2f_1^2 && g^2(f_1^2+f_2^2) && -g^2f_2^2 \\
0 && -g^2f_2^2 && \ddots
\end{array} \right).
\end{equation}

\vspace{0.1in} \noindent The mass eigenstates will have $\mathrm{mass} \sim gf_1,\ gf_2,\ gf_3,$ etc.  Thus, a given massive particle is identified with a given site.  We can integrate out particles simply by popping nodes off the UV end of the moose, then renormalizing $g$ and $f$, (Fortunately, for arguments based on the symmetry structure, we will not need the exact running of $g$.).  The locality maintained by holographic RG thus manifests itself as theory space locality.  Such a simple trick would not work in flat space.  The mass matrix in that setup would have a uniform value for $f$, and we could no longer affiliate a given mass eigenstate with a certain site.  

In the $\mathrm{U(1)}$ case, the link vevs Higgs the theory to the diagonal.  All link scalars are eaten in the process, and there is no candidate for a PGB.  Going now to MCHM, we give the leftmost site an $\mathrm{SU(2)\otimes U(1)}$ symmetry, the rightmost $\mathrm{SO(4)}$, and everything in between becomes $\mathrm{SO(5)}$.  

\begin{equation}
\begin{tabular}{c}
\xymatrix@R=.4pc@C=1.4pc{\mathrm{Global:} & SO(5) && SO(5) &  & SO(5) && SO(5) \\
& *=<20pt>[o][F]{} \doublerightxyarrow && *=<20pt>[o][F]{} \rightxyarrow & *=<20pt>[o]{\cdots} \rightxyarrow & *=<20pt>[o][F]{} \doublerightxyarrow && *=<20pt>[o][F]{} \\ \mathrm{Gauged:} & SU(2)\otimes U(1) && SO(5) &  & SO(5) & & SO(4)}
\label{eqn:5deconstruct}
\end{tabular}
\end{equation}

\vspace{0.15in} \noindent As before, ``integrating out" consists of removing nodes from the left side of the moose.  The complication in this setup comes from the mismatch between bulk and brane symmetries.  If we imagine starting at some high scale $\gg f_1$, we will have all gauge symmetries in the theory.  As we descend beneath $f_1$, the leftmost two gauge symmetries get Higgsed to their diagonal, which is $\mathrm{SU(2)\otimes U(1)}$.  Thus, to integrate out the first set of heavy modes, we remove the leftmost node, but reduce the symmetry on the adjacent one to $\mathrm{SU(2)\otimes U(1)}$.  Since the $\mathrm{SU(2)\otimes U(1)}$ site corresponds to the UV brane, we see that we have indeed moved it in toward the IR brane.  We can now take the extreme deconstructed limit, replacing the entire extra dimensional setup with a three-site moose.  The entire bulk of the extra dimension now consists of a single $\mathrm{SO(5)}$ site.  

We can determine the existence of a PGB in MCHM purely through Goldstone counting.  In limit that all gauge couplings turn off, the links have an $\mathrm{SO(5)^4}$ global symmetry, thus there will be 20 exact NGBs in the theory when they condense.  However, when the gauge groups break to the $\mathrm{SU(2)\otimes U(1)}$ diagonal, only 16 gauge bosons get masses, and we get a $\mathbf{4}$ of $\mathrm{SO(4)}$ pseudo.  Furthermore, its radiative potential is protected by collective symmetry breaking.  Setting to zero the gauge coupling on any particular site, we once again get 20 exact Goldstones.  With three gauge spurions needed to have a nonzero potential, we have recovered the finiteness of the Higgs mass.  Thus, the 5d warped model is protected by the same collective symmetry effect exploited in the little Higgs.  We can go further, reducing to two sites to recover the symmetry structure of the Custodial Minimal Moose\cite{Chang:2003un},\footnote{In the original setup, a full $\mathrm{SO(5)}$ was gauged, rather than an $\mathrm{SO(4)}$ subgroup.  This just has the effect of changing a quadruplet of heavy gauge bosons with no couplings to SM fermions into scalars.  These extra pseudos make no one-loop contributions to the dimension six operators constrained by observables.} which we will describe in detail below.  Working backward, we could have started with two sites, getting manifest locality in theory space.  By integrating in additional nodes with the appropriate warp factor, we can grow the extra dimension to an arbitrary number of sites while maintaining locality.  Any such nonlinear sigma model local in theory space is equivalent to Georgi's vector limit \cite{Georgi:1989xy}, and free from one loop (at least) quadratic divergences because of collective symmetry breaking \cite{Piai:2004yb}.

\section{The Custodial Minimal Moose}
\label{sec:cmm}

\hspace{0.15in} We will now review the setup of the Custodial Minimal Moose (CMM), a two-site, four-link little Higgs model \cite{Chang:2003un}.  In equation \ref{eqn:5deconstruct}, we only have a single extra dimension, and thus each site is connected to its neighbors by just one link. Deconstructing the theory to two sites, we cannot write down a nonderivative term only involving the link, $\Sigma$, without violating collective symmetry breaking.  Thus, the Higgs potential is purely radiative and we get no parametric separation between the electroweak and nonlinear sigma scales.  The solution for a two-site moose is to introduce three more links \cite{Arkani-Hamed:2002qx}. This allows us to write plaquette operators that can generate an $\mathcal{O}(1)$ quartic.  We give the details in the next subsection. The links transform under global $\mathrm{SO(5)_L\otimes SO(5)_R}$ symmetries with gauged $\mathrm{SU(2)\otimes U(1)}$ and $\mathrm{SO(4)}$ subgroups, respectively.  

\begin{equation}
\begin{tabular}{c}
\xymatrix@R=.4pc@C=1.4pc{SO(5) && SO(5)  \\
*=<20pt>[o][F]{} \doublerightDDxyarrow \doublerightDxyarrow \doublerightUxyarrow \doublerightUUxyarrow && *=<20pt>[o][F]{} \\ 
SU(2)\otimes U(1) && SO(4) }
\end{tabular}
\end{equation}

\vspace{0.15in} \noindent The nonlinear sigma fields Higgs these groups to the diagonal.  The advantage of this setup over the original Minimal Moose \cite{Arkani-Hamed:2002qx} is the existence of an explicit custodial symmetry, $\mathrm{SU(2)_C}$.  Contributions to the $T$ parameter from the heavy gauge boson $B'$ are partially canceled by the charged bosons in its $\mathrm{SU(2)}$ multiplet.  This cancellation allows us to take a limit in which $W'$ and $B'$ decouple from SM fermions, thus suppressing contributions to observables from 4-fermi and Higgs-fermi operators.  Our motivations for choosing an $\mathrm{SO(5)}$ global symmetry are the same as in MCHM.  It is the smallest example of a group which contains both $\mathrm{SU(2)_L\otimes SU(2)_R}$ as a subgroup and a field in its adjoint with the right quantum numbers for the Higgs.  We will also consider versions of the model, which differ from that in \cite{Chang:2003un}.  The third generation quarks can be modified to implement the left-right parity proposed in \cite{Agashe:2006at} to protect $Z\rightarrow\overline{b}_Lb_L$.  This provides an example of translating features from 5d to a little Higgs model.  Additionally, we will discuss the idea of \cite{Arkani-Hamed:2002pa, Birkedal-Hansen:2003mp}, imposing an exact $\mathbb{Z}_2$ symmetry that allows for stable dark matter candidates.

\subsection{Gauge and Scalar Sectors}
\label{sec:gaugescalar}

\hspace{0.15in} The link fields acquire vevs proportional to the identity at a scale $f$, breaking the gauge symmetry to $\mathrm{(SU(2)\otimes U(1))_V}$, giving masses to six gauge bosons.  An approximate $\mathbb{Z}_4$ symmetry among the links ensures that symmetry breaking occurs for each one at the same scale. We can study interactions in the theory by expanding in linearized fluctuations, $\pi$, about the vacuum.

\begin{equation}
\Sigma = \exp\left( i\frac{\pi^a t^a}{f}\right),
\end{equation}

\noindent where the $t^a$ are $\mathrm{SO(5)}$ generators in the vector representation, normalized as in \cite{Agashe:2004rs}. Since the linear combination $\pi_1+\pi_2+\pi_3+\pi_4$ shifts under the broken generators, we can identify the $\mathbf{6}$ of $\mathrm{SO(4)}$ components as the eaten Goldstones.  This leaves 34 uneaten scalars that acquire radiative potentials \cite{Coleman:1973jx}.  To determine the masses of the heavy gauge bosons, we start with the lagrangian for a gauged nonlinear sigma model:

\begin{eqnarray}
\mathcal{L} &=& \frac{1}{2}f^2 \mathrm{Tr} (D_\mu \Sigma_i)^\dagger D^\mu \Sigma_i \\
D^\mu \Sigma_i &=& \partial_\mu \Sigma_i-i(g_2 A_2^\mu+g_1 A_1^\mu)\Sigma_i+ig_4 \Sigma_i A^\mu_4.
\end{eqnarray}

\noindent At lowest order in the pion expansion, we get a mass matrix for the gauge bosons, and the following eigenstates:

\begin{table}[ht]
\begin{center}
\begin{tabular}{|c|c|c|}
\hline Mass & Content & Name \\\hline\hline
 & & \\
0 & $\cos\theta'\: B_1+\sin\theta'\: W_4^{r3}$ & $B$ \\\hline
 & & \\
0 & $\cos\theta\: W_2^a+\sin\theta\: W_4^{\ell a}$ & $W^a$ \\\hline
 & & \\
$\frac{16g'^2f^2}{\sin^2 2\theta'}$ & $\cos\theta'\: W_4^{r3}-\sin\theta'\: B_1$ & $B'$ \\\hline
 & & \\
$\frac{16g^2f^2}{\sin^2 2\theta}$ & $\cos\theta\: W_4^{\ell a}-\sin\theta'\: W_2^a$ & $W'^a$ \\\hline
 & & \\
$\frac{16g'^2f^2\cos^2\theta'}{\sin^2 2\theta'}$ & $W^{r\pm} $ & \\\hline
\end{tabular}
\end{center}
\caption{Gauge Bosons in the Custodial Minimal Moose}
\end{table}

\noindent where,

\begin{eqnarray}
\nonumber
g'=\frac{g_1 g_4}{\sqrt{g_1^2+g_4^2}} &\ \ \ & g=\frac{g_2 g_4}{\sqrt{g_2^2+g_4^2}} \\
\nonumber
\cos\theta'=g'/g_1 &\ \ \ & \cos\theta=g/g_2 \\
\sin\theta'=g'/g_4 &\ \ \ & \sin\theta=g/g_4.
\end{eqnarray}

The cutoff of the nonlinear sigma theory will be $\mathcal{O}(4\pi f)$.  However, the little Higgses are protected from getting a mass at this scale by collective symmetry breaking.  Turning on both couplings, there is only a log divergent mass above the scale of heavy gauge bosons.  Beneath this scale, we generate $m_{\mathrm{scalar}}^2\sim \frac{g^4 f^2}{16\pi^2}$, where $g^4$ is some linear combination of the three gauge couplings to the fourth.  Unfortunately, the quartic for the theory is also suppressed by a gauge spurion, $\sim\frac{g^2}{16\pi^2}$.  Without an $\mathcal{O}(1)$ quartic, the electroweak breaking scale, $v$, and the nonlinear sigma breaking scale, $f$, will slide together.  In the warped setups, the solutions proposed are to tune the Coleman-Weinberg potential directly to increase the quartic, or, as discussed above, add new tree level terms and tune them against each other to cancel the mass contribution.  In the little Higgs, we can add tree level quartics without generating mass terms or using tuning.  One can understand this difference by the dimensions involved.  In MCHM, the Higgs is the zero mode of the bulk $A^5$ field.  Thus, any potential generated will necessarily involve the symmetry breaking terms on the boundaries.  $A\  priori$, there is no reason to generate a quartic without a quadratic term.  In a little Higgs setup, however, we can imagine discretizing additional dimensions.  One can think of the Minimal Moose as discretized 2-torus with periodic identifications, and a $\mathbb{Z}_2$ orbifolding \cite{Gregoire:2002ra}.  Since it descends from a 6d theory, there is automatically a notion of quartic without quadratic, the commutator potential of the gauge kinetic term, $\left[A^5,A^6\right]^2$ \cite{Arkani-Hamed:2001nc}.  

In a discretized theory, such a potential takes the form of a plaquette operator.  Once again, we can understand the absence of a mass term by collective symmetry breaking.  We want plaquettes such that one combination of pions get a mass, and integrating them out generates a quartic for the little Higgses.  One possibility is:

\begin{equation}
\mathcal{L}_{plaq} = \kappa_1 f^4 \mathrm{Tr}\left(\Sigma_1\Sigma_2^\dagger\Sigma_3\Sigma_4^\dagger\right)+\kappa_2 f^4 \mathrm{Tr}\left(\Sigma_1\Sigma_4^\dagger\Sigma_3\Sigma_2^\dagger\right)+\mathrm{h.c.}
\label{eqn:plaq}
\end{equation}

\noindent This gives a mass to the linear combination, $\pi_1-\pi_2+\pi_3-\pi_4$.  The individual plaquettes each preserve an $\mathrm{SO(5)}^4$ symmetry, leaving the other three linear combinations as exact NGBs after symmetry breaking.  Together, however, the $\kappa_1$ and $\kappa_2$ terms only protect the eaten one.  No mass term is generated because there are no linear couplings between heavy and light fields.  In Table \ref{tbl:xxx}, we list the orthogonal combinations of pions that will serve as the exact Goldstones, quartic-generating heavy fields, and little Higgses.  \nolinebreak
\begin{table}[ht]
\begin{center}
\begin{tabular}{|c|c|c|}
\hline Name & Content & Characteristic \\\hline\hline
$w$ & $\frac{1}{2}(\pi_1+\pi_2+\pi_3+\pi_4)$ & $\mathbf{6}$ eaten \\
  & & $\mathbf{4}$ light \\\hline
$z$ & $\frac{1}{2}(\pi_1-\pi_2+\pi_3-\pi_4)$ & heavy \\\hline
$x$ & $\frac{1}{\sqrt{2}}(\pi_1-\pi_3)$ & light \\\hline
$y$ & $\frac{1}{\sqrt{2}}(\pi_2-\pi_4)$ & light \\\hline
\end{tabular}
\end{center}
\caption{Pion decuplet linear combinations}
\label{tbl:xxx}
\end{table}
\noindent With these combinations of decuplets, after integrating out the $z$, we get:

\begin{equation}
\mathcal{L}_{plaq} = \frac{c_1 c_2}{2(c_1+c_2)}\mathrm{Tr}([x,y]^2),
\end{equation}

\noindent where $c_i=2\mathrm{Re}\;\kappa_i$.  The approximate $\mathbb{Z}_4$ symmetry on the links sets $c_1$ and $c_2$ nearly equal.  For proper electroweak symmetry breaking, we will need to add further plaquette terms.  As they break additional symmetries, we can expect their contributions to be suppressed by a loop factor compared to equation \ref{eqn:plaq}.  The deformation necessary for EWSB is:

\begin{eqnarray}
V_{\epsilon Plaq}=-i\epsilon f^4 \mathrm{Tr}\;T^{r3}\left(\Sigma_1\Sigma_2^\dagger\Sigma_3\Sigma_4^\dagger+\Sigma_1\Sigma_4^\dagger\Sigma_3\Sigma_2^\dagger\right)+\mathrm{h.c.} \\
V_{\epsilon Plaq}\supset 4i\epsilon f^2 \mathrm{Tr}\;T^{r3}[x,y]. 
\label{eqn:eplaq}
\end{eqnarray}

Finally, we may wish to decouple the scalars that transform as $\mathbf{6}$s of SO(4).  A light scalar triplet under SU(2) can get a vev from the interaction $h^\dagger\phi h$, leading to a large contribution to $\delta\rho$.  However, we can raise the $\phi$ masses by using $\Omega$ plaquettes as in \cite{Chang:2003un,Gregoire:2002ra}, with $\Omega=\exp(2\pi i t^3_R)=\mathrm{Diag}(-1,-1,-1,-1,1)$.  For example, to raise the $\mathbf{6}$ in the $x$ combination of pions, we can use:

\begin{equation}
V_{\Omega Plaq}=\lambda_\Omega f^2 \mathrm{Tr}\left(\Omega\Sigma_1\Sigma_3^\dagger\Omega\Sigma_1\Sigma_3^\dagger\right)+\mathrm{h.c.}
\end{equation}  

\subsection{Fermion Sector}

\subsubsection{Original CMM}

\hspace{0.15in} Here we present the fermion sector as originally written down in \cite{Chang:2003un}.  Light fermions are allowed to contribute to quadratically divergent Higgs mass terms.  Their small yukawas make such contributions sub-leading for $\Lambda\sim10$ TeV.  We can couple our fermions to the link fields by embedding them in $\mathbf{4}$s of SO(5).  Similarly, we can change the link field from a bi-fundamental to a bi-spinor:

\begin{equation}
\slashed{\Sigma}_i \equiv \exp^{i\pi^a\tilde{T}^a},
\end{equation}

\noindent where the $\tilde{T}^a$ are the generators of the spinor representation of SO(5).  Under the SO(5) global symmetries, $\slashed{\Sigma}_i$ transforms as $\slashed{\Sigma}_i\rightarrow L_i \slashed{\Sigma}_i R_i^\dag$, where $L$ and $R$ are spinor transformations.  This allows for a simple coupling of SM fermions.  To avoid FCNCs, we will couple all fermions (including $3^\mathrm{rd}$ generation quarks) to one linear combination of pions, which we choose to be $x=\frac{1}{\sqrt{2}}\left(\pi_1-\pi_3\right)$.  The mass terms for up type quarks are given by:

\begin{eqnarray}
\nonumber
\mathcal{L}_{\mathrm{up\ quarks}} &=& \mathcal{U}^{c\top} \slashed{\Sigma}_1\slashed{\Sigma}^\dagger_3\mathcal{Q} +\ \mathrm{h.c.}\\
\nonumber
\mathcal{Q}^{\top}&=&(u,d,0,0) \\
\mathcal{U}^{c\top}&=&(0,0,u^c,0),
\label{eqn:lfyukawa}
\end{eqnarray}

\noindent with similar terms for down-type quarks and leptons.  For third generation quarks, we will need to add new fields to implement collective breaking.  Naively, one can add a  right-handed doublet to $q_3$, giving $(q_3^\top,\tilde{u},\tilde{d})$, and then marry off $\tilde{u},\tilde{d}$ to $\tilde{u}^c,\tilde{d}^c$.  However, taking advantage of an accidental SU(3) symmetry,\footnote{This is the SU(3) subgroup of the SU(4) that acts on the links and fermion vectors.  In fact, it is only approximate, but exact to quadratic order in the Higgs field \cite{Chang:2003un}.} one can decouple the $\tilde{d}$ and have $\mathcal{Q}_3=(q_3^\top,\tilde{u},0)$.  The following lagrangian will thus have no quadratically divergent Higgs mass:

\begin{equation}
\mathcal{L_\mathrm{top}}=y_1 f\mathcal{U}^{c\top}_3\slashed{\Sigma}_1\slashed{\Sigma}^\dagger_3\mathcal{Q}_3 +y_2 f\tilde{u}\tilde{u}^c+\ \mathrm{h.c.}
\end{equation}   

\noindent The $u^c_3-\tilde{u}$ mixing gives a physical top quark and a new $t'$.  The top yukawa is given by:

\begin{equation} 
y_{\mathrm{top}}=\frac{|y_1||y_2|}{\sqrt{2(|y_1|^2+|y_2|^2)}},
\end{equation}

\noindent and the $t-t'$ mixing angle $\theta_y$ is defined as $\tan\theta_y=|y_1|/|y_2|$.

\subsubsection{$P_{LR}$ Invariant third generation}

\hspace{0.15in} One of the original difficulties in MCHM was the presence of a large correction to $Z\rightarrow \overline{b}_L b_L$ \cite{Agashe:2005dk}.  To remedy this, the authors of \cite{Agashe:2006at} suggested using a parity that exchanges the $\mathrm{SU(2)_L}$ and $\mathrm{SU(2)_R}$ involved with custodial $\mathrm{SU(2)}$ \cite{Sikivie:1980hm}.  It thus enhances $\mathrm{SU(2)_L}\otimes\mathrm{SU(2)_R}\cong SO(4)$ to O(4).  The coupling of the $Z$ boson to left-handed SM fermions, $\psi$, is given at zero momentum by:

\begin{equation}   
\frac{g}{\cos \theta_W}\left[Q^3_L-Q \sin^2 \theta_W\right]Z^\mu\overline{\psi}\overline{\sigma}_\mu\psi.
\end{equation}

\noindent If the operator that corrects $Q^3_L$ is a linear coupling between $\psi$ and some BSM operator $\mathcal{O}_\psi$, it can only be invariant under $P_{LR}$ if $\psi$ is a $P_{LR}$ eigenstate.  In our setup the BSM fields will consist of links and gauge bosons.  Neither of these couplings are linear, but both still require $\psi$ to be a $P_{LR}$ eigenstate for invariance.  Furthermore, we can demand that the interaction respect $\mathrm{SU(2)_L\otimes SU(2)_R}$ since our BSM sector is invariant under O(4).  In that case, $\psi$ will take the appropriate $T_L,T_L^3,\ etc.$ quantum numbers.  Specifically, we will want to maintain an invariance under $\mathrm{U(1)_V}$, so that $\delta Q_V=\delta Q_L+\delta Q_R=0$.  If $\psi$ is a $P_{LR}$ eigenstate, then it must have $T_L=T_R$ and $T^3_L=T^3_R$.  Therefore, $\delta Q_L=\delta Q_R$, and thus $\delta Q_L=0$.  Of course, in a standard embedding, as in the CMM, $b_L$ does not have the same charge under $\mathrm{SU(2)_L}$ as $\mathrm{SU(2)_R}$, as it is in a $(\mathbf{2},\mathbf{0})$ representation.  Thus, to take advantage of the discrete symmetry protection, we need a different third generation.  

\begin{figure}[h!]
\begin{center}
\includegraphics{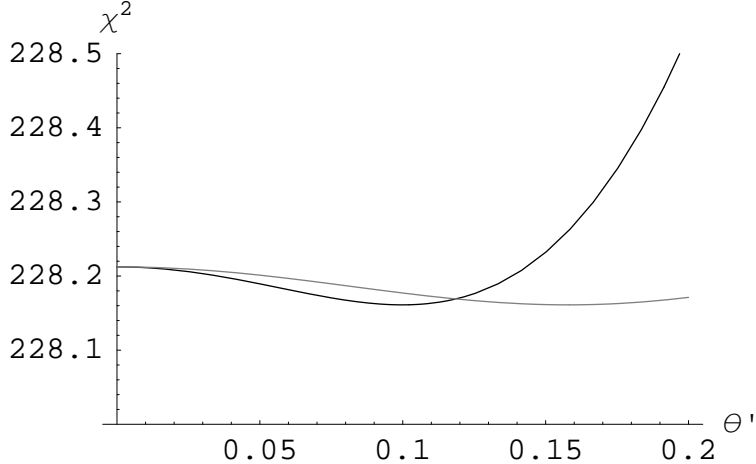}
\end{center}
\caption{For original CMM (lighter curve) and CMM with modified third generation (darker), we calculate the $\chi^2$ compared to data as a function of $\theta'$ and $f$=700 GeV for  predictions of operator coefficients $a_{hQ}^t$ and $a_{hQ}^s$, using the method of \cite{Han:2004az,Han:2005pr} (See Appendix for details).  $\chi^2_\mathrm{min}=228.1$.}
\label{fig:zchi2}
\end{figure}

As we will see, even though no attempt was made to suppress $\delta g_{Lb}$ specifically in \cite{Chang:2003un}, the ability to take a limit in which all nonoblique corrections vanish works just as well ($cf.$ Figure \ref{fig:zchi2}).  However, we include this alternate possibility as an example of taking a feature from a 5d warped model and putting into 4d.  It also shows once again how one can obtain the same end result, but with less up-front labor in four dimensions.  In the $\mathrm{MCHM_5}$ of \cite{Contino:2006qr}, the third generation was embedded in a $(\mathbf{2},\mathbf{2})$ of $\mathrm{SU(2)_L\otimes SU(2)_R}$ and placed on the IR brane.  The extra SU(2) doublet is projected out by Dirichlet boundary conditions at the UV brane.  We will write the new third generation multiplet as  
 
\begin{equation}  
Q_L = Q_L^\alpha \sigma^\alpha = \left( \begin{array}{cc}
t_L & C_{5/3L} \\
b_L & C_{2/3L} \end{array} \right),
\end{equation}

\noindent where $\sigma^\alpha = (\mathbf{1},i\sigma_1,i\sigma_2,i\sigma_3)$.  After \cite{Contino:2006qr}, we call the extra doublet the $custodians$, and label them by their electric charges.  The analogue of the IR brane is the SO(4) site, and that is where we need to place $Q_L$ to take advantage of the explicit $O(4)$.  In order to give $q_{L}^3\equiv(t_L,b_L)$ the right $\mathrm{U(1)_Y}$ charge, we will delocalize the $U(1)_X$ of the $\mathrm{SU(2)\otimes U(1)}$ site and give $Q_L$ a charge of 2/3 under it.  The spontaneous breaking of the links to the diagonal sets $Y=X+t^3_R$. There is also a more manifestly SO(4) way to write $Q_L$:

\begin{equation}
Q_L = \left( \begin{array}{c}

-i(C_{5/3}+b) \\
(C_{5/3}-b) \\
i(C_{2/3}-t) \\
(t+C_{2/3}) \\\end{array} \right)/\sqrt{2}.
\end{equation}

\noindent In order to maintain collective symmetry breaking, we need to keep a manifest SO(5) symmetry in the coupling to the link fields.  Thus, we will complete $Q_L$ into a $\mathbf{5}$ of SO(5), $\hat{Q}_L^\top=(Q_L, T)$.  Since $T$ is an SO(4) singlet, we can give it a mass by coupling it to a $T^c$ living on the SO(4) site.  Getting rid of the $C$ fields is not as trivial.  The 4d analogue of Dirichlet boundary conditions on the UV brane is a link operator connecting $Q_L$ to a doublet on the SU(2) site, $C_L^c$.  We can embed it in an SO(4) vector:

\begin{equation} 
C_L^c = \left( \begin{array}{c}

iC_{5/3}^c \\
C_{5/3}^c \\
-iC_{2/3}^c \\
C_{2/3}^c \end{array} \right).
\end{equation}

\noindent The potential for the top sector is thus:

\begin{equation}
V_{t} = \lambda_1 f C_L^{c\top} \Sigma_1 \hat{Q}_L+\lambda_2 f \hat{Q}_L^\dag\Sigma_3^\dagger\Sigma_1 \left(\begin{array}{c}\mathbf{0}\\
u_3^c \end{array} \right)+\lambda_3 f TT^c+\ \mathrm{h.c.}
\end{equation}

\noindent The right handed top will be a linear combination of the singlets, $C_{2/3}^c,\ u_3^c,$ and $T^c$.  We see that the top/Higgs term thus has an $\mathrm{SO(5)^6\times SO(4)}$ symmetry, which leaves enough symmetry to protect all scalars.  Only by including it along with the $\lambda_1$ or $\lambda_3$ operators can one generate a mass for the Higgs.  Although the bottom is allowed to violate collective symmetry breaking, the embedding of $q_3^L$ into $Q_L$ requires a somewhat unconventional term if we do not wish to add more fields:

\begin{eqnarray}
V_b &=& \mathrm{Tr}\left[\slashed{\Sigma}_1 \left( \begin{array}{cc}
0 & Q_L \\
0 & 0 \end{array} \right) \left( \begin{array}{cc}
0 & 0 \\
0 & B \end{array} \right)\slashed{\Sigma}_3^\top \right] \\
B &=& \left( \begin{array}{cc}
0 & b^c \\
0 & 0 \end{array} \right).
\end{eqnarray}

We can now check how this modification does in comparison to the original CMM.  Changes to $g_{Lb}$ come from the dimension six operators $\mathcal{O}_{hQ}^s$ and $\mathcal{O}_{hQ}^t$.  These couple two Higgs fields to two third generation quarks.  They arise from integrating out $B'$ and $W'$ at tree level, respectively.  Whatever the third generation, as long as it only couples to the $x$ linear combination of pions, we get the following correction to $Z\rightarrow\overline{b}_L b_L$:

\begin{equation}
\delta g_{Lb}=\sqrt{g^2+g'^2}\;v^2\cos^2 \beta\left(a_{hQ}^t+a_{hQ}^s\right),
\label{eqn:deltagl}
\end{equation}

\noindent where $\langle h_x \rangle =(0,\; v\cos(\beta)/\sqrt{2})$.  The $a$ coefficients are given for the original CMM in equation \ref{eqn:dim6}.  For the modified third generation, they are:

\begin{eqnarray}
a_{hQ}^s = \frac{\cos2\theta'\sin^2\theta'}{24f^2} + \frac{\cos2\theta'\cos^2\theta'}{32f^2}&&a_{hQ}^t =-\frac{\cos2\theta\cos^2\theta}{32f^2}. 
\end{eqnarray}

\begin{figure}[h!]
\begin{center}
\includegraphics{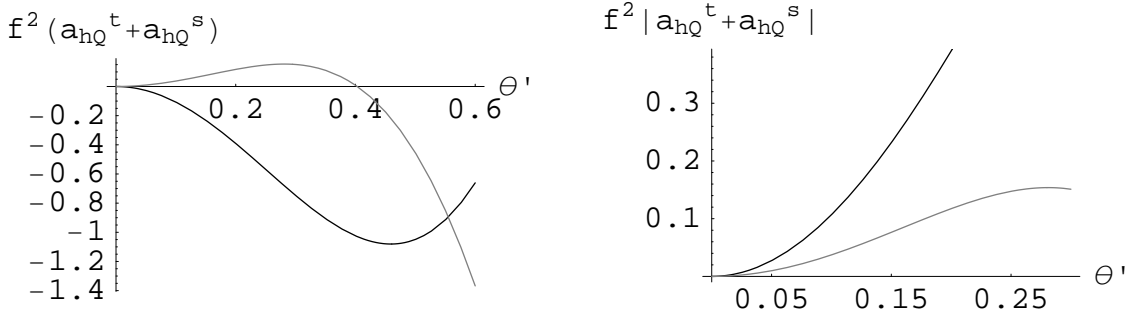}
\end{center}
\caption{For original CMM (lighter curve) and CMM with modified third generation (darker) L: $a_{hQ}^t+a_{hQ}^s$ vs. $\theta'$ and R: $|a_{hQ}^t+a_{hQ}^s|$ vs. $\theta'$}
\label{fig:ahQ3ahQ1}
\end{figure}

\noindent The first term in $a_{hQ}^s$ comes from the $\mathrm{U(1)_X}$ charge of $b_L$.  The second term in $a_{hQ}^s$ and $a_{hQ}^t$ are from the $b_L$ coupling to SO(4).  Note that they cancel in the limit that $\theta\rightarrow\theta'$, when $P_{LR}$ is restored as a symmetry.  To compare the two cases, we divide out by the common factor in equation \ref{eqn:deltagl} and plot $f^2(a_{hQ}^s+a_{hQ}^t)$ as a function of $\theta'$ (Figure \ref{fig:ahQ3ahQ1}). \nolinebreak
\noindent  In fact, the original setup fares better except at large $\theta'$.  Taking the limit $\theta'\rightarrow\frac{\pi}{2}$ would allow us to use both angular suppression and approximate symmetries for cancellation.  This is the limit in which the SM gauge bosons become those of the SO(4) gauge group.  The value of $\theta_w\approx\frac{\pi}{6}$ and the fact that $\sin\theta=\cot\theta_w \sin\theta'$ means that we cannot take $\theta'>0.62$.  Figure \ref{fig:ahQ3ahQ1} shows that even in taking $\theta'$ as large as possible, the value of $\delta g_{Lb}$ in the modified theory is still several times that for small $\theta'$ in the original setup.  

The modified version induces a negative correction to the coupling, which is less constrained \cite{Agashe:2005dk}.  At small $\theta'$, the original setup has $\delta g_{Lb}>0$.  Nonetheless, comparisons to bounds on $\mathcal{O}_{hQ}^s$ and $\mathcal{O}_{hQ}^t$ show the two models perform equally well for $\theta'\lesssim 0.12$, after which the original fares better (Figure \ref{fig:zchi2}). \nolinebreak
\noindent Thus, the original model already suppressed $Z\rightarrow\overline{b}_L b_L$ well within its experimental limits.  Modifying the third generation causes no improvement.  Given this, for subsequent $\chi^2$ analysis and comparisons to 5d warped model, we will retain the original fermion sector.

\subsection{Electroweak Symmetry Breaking}

\hspace{0.15in} As mentioned in Section \ref{sec:gaugescalar}, viable EWSB requires the presence of an ``$\epsilon$-plaquette," one which breaks more global symmetries than those responsible for the Higgs quartic.  Thus, we expect $\epsilon\sim 0.01\lambda$.  The tree-level and one-loop Coleman-Weinberg Higgs potential is the following:

\begin{equation}
V_\mathrm{Higgs} \simeq \frac{\lambda}{2}(|h_x^\dag h_y-h_y^\dag h_x|^2+4|h_x^a h_y^b \epsilon^{ab}|^2)+(ib\,h_x^\dag h_y + h.c.)+m_x^2|h_x|^2+m_y^2|h_y|^2.
\label{eqn:Higgspotl}
\end{equation}

\noindent We have only included the plaquette contribution to the quartic as that is the dominant one $i.e.\ \lambda=\frac{c_1 c_2}{2(c_1+c_2)}$ as in Section \ref{sec:gaugescalar}.  Also, $b=2\epsilon f^2$.  The gauge and scalar Coleman-Weinberg potentials contribute to both $m_x^2$ and $m_y^2$, but the fermions only to $m_x^2$.  

Minimizing the potential away from the origin gives the following vevs:

\begin{equation}\label{eq:vev}
\langle h_x \rangle = \left(\begin{array}{c}
0 \\
\frac{v \cos\beta}{\sqrt{2}} \end{array} \right),\ \ \langle h_y \rangle = \left(\begin{array}{c}
0 \\
\frac{v \sin\beta e^{i\phi}}{\sqrt{2}} \end{array} \right).
\end{equation}

\noindent The angle $\phi$ is defined by $\cos\phi=m_{xy}^2/m_x m_y$, where $m_{xy}^2\,h_x^\dag h_y$ + h.c. is a term absent in the potential to the order we are considering.  Thus, we will have $\phi\approx\frac{\pi}{2}$, and our vacuum maximally breaks custodial SU(2).   To avoid an unbounded from below potential, we must impose the following conditions:

\begin{eqnarray}
\nonumber
m_x^2,\ m_y^2>0\\
m_x^2\, m_y^2 - m_{xy}^4 &>& 0.
\end{eqnarray}

\noindent These prevent problems with $h_y=0$, $h_x=0$, and $\phi=0$ flat directions, respectively.  Additionally, having $h_x=h_y=0$ as an unstable point requires

\begin{equation}
m_x^2\, m_y^2 - m_{xy}^4 -b^2 < 0.
\end{equation}

\noindent In the case that $\phi$ is exactly $\frac{\pi}{2}$, the physical Higgses fall into $CP$ eigenstates, and we have the same five scalars as in the MSSM.  The Appendix has the complete listing of physical Higgs parameters in terms of those in equation \ref{eqn:Higgspotl}.

\subsection{Dark Matter Variant}
\label{sec:dmv}

\hspace{0.15in} We will briefly discuss how one can implement a dark matter candidate within the CMM.  In \cite{Arkani-Hamed:2002pa}, the authors discussed how to make the discrete, geometric symmetries of theory space exact even after EWSB.  This allows for a lightest parity odd particle and dark matter candidate.  The presence and cosmological viability of such a candidate in the CMM was demonstrated in \cite{Birkedal-Hansen:2003mp}.  We present here a slightly different version, one that requires the addition of less than half as many quarks as the original.  As mentioned above, there is an approximate $\mathbb{Z}_4$ symmetry under which the links are rotated into each other.  It is the remnant of 6d rotational symmetry.  We start by taking the $\mathbb{Z}_2$ subgroup, where $\Sigma_x(\equiv \Sigma_1\Sigma_3^\dag)\rightarrow\Sigma_x^\dag$, since $1\rightarrow 3\rightarrow 1$ under the transformation.  Thus, we can no longer couple the fermions to the Higgs as in equation \ref{eqn:lfyukawa}.  However, it is straightforward to make an invariant interaction.  We can use:

\begin{equation}
\mathcal{L}_{\mathrm{up\ quarks}} = \lambda_q f \mathcal{U}^{c\top} (\slashed{\Sigma}_x-\slashed{\Sigma}_x^\dag)\mathcal{Q} +\ \mathrm{h.c.},
\label{eqn:dmlfyukawa}
\end{equation}

\vspace{0.1in} \noindent provided that $\mathcal{U}^c\mathcal{Q}\rightarrow -\mathcal{U}^c\mathcal{Q}$ under the link symmetry, which we will call $R_2$, since it is not a $\mathbb{Z}_2$ when acting on the fermions.  Unfortunately, this term violates collective symmetry breaking, and thus we cannot use it for the third generation.  For this, we add four $\left(\mathbf{2},\mathbf{1}\right)\oplus\left(\mathbf{1},\mathbf{2}\right)$ $\chi$ quarks to the SO(4) site, and couple them to the third generation:

\begin{equation}
\mathcal{L}_{\mathrm{DM\ top}} = -i\left[M_L\mathcal{Q}_3 \left(\Sigma_3\chi^c + i \Sigma_1\tilde{\chi}^c \right) + M_R\left(\chi \Sigma_1^\dag + i \tilde{\chi}\Sigma_3^\dag \right)u_3^c \right]-M\left(\chi \chi^c+ \tilde{\chi} \tilde{\chi}^c \right).
\end{equation}

\noindent Upon integrating out the $\chi$, we get a term identical to equation \ref{eqn:dmlfyukawa}, with $\lambda_Q=\frac{M_L M_R}{M}$.  Under $R_2$, the third generation must transform as follows:

\begin{eqnarray}
\nonumber
\mathcal{Q}_3 &\rightarrow& i\mathcal{Q}_3 \rightarrow -\mathcal{Q}_3 \\
\nonumber
u_3^c\ &\rightarrow& i u_3^c \rightarrow -u_3^c \\
\nonumber
\chi &\rightarrow& \tilde{\chi} \rightarrow -\chi \\
\chi^c &\rightarrow& \tilde{\chi}^c \rightarrow -\chi^c.
\end{eqnarray}

\noindent Clearly, after EWSB, $R_2$ is no longer a good symmetry of theory.  However, we can use $\Omega=\exp\left(2\pi i \tilde{T}^3_R\right)=\mathrm{Diag}\left(1,1,-1,-1\right)$ in the spinor representation.  It causes $\mathrm{SU(2)}_L$ singlet fermions and the $\mathrm{SU(2)}_L$ doublet scalars to pick up a minus sign.  Thus, composing this $\mathbb{Z}_{2\Omega}$ with $R_2$, we get an exact symmetry even after EWSB.  Since the $\mathrm{SU(2)}_L$ triplet scalars do not pick up a sign under $\mathbb{Z}_{2\Omega}$, they will be parity odd, and thus the lightest one can serve as the dark matter.  One might worry that adding four new $\mathrm{SU(2)}_L$ doublets will cause problems with the $S$ parameter.  Even if this contribution were 10 times its naive estimate, we get acceptable contributions for $M\gtrsim 5$ TeV.  At this scale, we still are not forced into tuning the Higgs potential.  

\section{Precision Electroweak}
\label{sec:PEWk}

\hspace{0.15in} Both the Custodial Minimal Moose and the 5d Composite Higgs have built in symmetries to minimize their tensions with experimental constraints.  In the former, we have an explicit custodial $\mathrm{SU(2)}$, which minimizes contributions to the $T$ parameter and allows for decoupling nonoblique corrections.  In addition to $\mathrm{SU(2)_C}$, the 5d Composite Higgs also has a LR parity to suppress the leading nonoblique correction, $Z\rightarrow \overline{b}_L b_L$.  In \cite{Carena:2007ua}, a global fit on a particular version of the 5d Composite was done using the method of \cite{Han:2004az,Han:2005pr}.  We perform here the same on the CMM.  Before we present these results, we will review the tensions within the 5d model.  

\subsection{5d Composite Higgs in a Corner}

\hspace{0.15in} Like any large $N$ technicolor model, the 5d composite Higgs has a sizable, positive contribution to the $S$ parameter.  In this case,

\begin{eqnarray}   
S=\frac{3}{8}\frac{N}{\pi}\epsilon^2 \\
\frac{1}{N} \equiv \frac{g_5^2 k}{16\pi^2},
\end{eqnarray}

\noindent where $\epsilon=\frac{v}{f}$, and $f$ is the decay constant of the nonlinear sigma field in which the Higgs lives.  Even for a ``large" $N$ of 5, satisfying the $S$ constraint still forces 10\% tuning of the terms of the Higgs potential against each other.  What is more, having an acceptable $S\ (\approx 0.3)$ requires a positive $T$ parameter of roughly the same size.  In \cite{Carena:2006bn,Carena:2007ua}, the authors perform numerical calculations for the $T$ parameter.  In their setup, quarks live in $\mathbf{5}$s of SO(5), with LH doublets in a $\left(\mathbf{2},\mathbf{2}\right)$ of \custodial in order to cancel contributions to $\delta g_{Lb}$.  There are two competing effects.  The additonal SU(2) doublet paired with $q^3_L$ gives a negative contribution to $T$.  There is, however, a positive contribution from the SO(4) singlet that completes the representation.  

For viable EWSB and $m_\mathrm{top}$, the top quark localizations are constrained, $|c_Q,c_u|<0.5$.  Unfortunately, for the allowed values of $c_Q$, positive values of $T$ only occur for $c_U$ (the $t_R$ localization) $\lesssim$ -0.4.  Furthermore, the top KK modes also induce one-loop corrections to $g_{Lb}$.  Thus, $T$ cannot be made too large ($\gtrsim$0.1) without pushing the $Z\rightarrow\overline{b}_L b_L$ rate 2$\sigma$ from its observed value.  Getting an acceptable value of $S$ for this value of $T$ will require still more tuning in the Higgs potential.  Requiring $c_U$ to take values near -0.4 is problematic as the top Yukawa vanishes in the limit that it is taken to -0.5.  $M_U$ is the mixing mass on the IR brane responsible for generating the top mass.  It couples the SO(4) singlet corresponding to $t_R$ to the SO(4) singlet living in the same $\mathbf{5}$ of SO(5) as $t_L$.  Ref. \cite{Carena:2006bn} gets:

\begin{eqnarray}
\nonumber
\lambda_{\mathrm{top}}\approx g M_U\left[\frac{(\frac{1}{2}-c_Q)(\frac{1}{2}+c_U)2kLe^{2(1+c_U-c_Q)}}{(1-e^{(1-2c_Q)kL})(1-e^{(1+2c_U)kL})} \right]^{1/2} \\
\times\left[1+M_U^2 e^{2(c_U-c_Q)kL}\frac{(\frac{1}{2}+c_U)(1-e^{(1-2c_Q)kL})}{(\frac{1}{2}+c_Q)(1-e^{(1+2c_U)kL})}\right]^{-1/2}.
\end{eqnarray}              

\noindent As usual, $k$ is the AdS curvature; $L$ is the position of the TeV brane.  In order to compensate the effect of sending $c_U\rightarrow-0.5,$ we will need to increase $M_U$.  However, this will raise the scale of the KK modes giving positive contributions to the $T$ parameter, thus suppressing them.  

The right numeric values for $S$ and $T$ require tuning.  Unfortunately, there does not seem to be a parameter range where both tunings are moderate $\sim10\%$.  The ultimate physical origin of the problem lies in the fact that we are both large $N$ and custodial symmetry respecting.  One parameter is naturally large and the other naturally small.  An additional tension is introduced when we go to protect $Z\rightarrow\overline{b}_L b_L$.  The fields necessary to do this make the situation worse, driving $T$ negative.  Restoring a positive contribution restricts $c_U$ to $<10\%$ of the range it is naively allowed by EWSB.  This range is further cut by corrections to $\delta g_{Lb}$ and by the size of the top yukawa.  The latter constraint could be significant as a large top yukawa is in direct conflict with a postive $\Delta T$.

\subsection{Constraints for CMM}
\label{subsec:PEWkCMM}

\hspace{0.15in} To test the CMM, we compare the predicted dimension six operators to the data via the following $\chi^2$ function:

\begin{equation}
\chi^2=\chi^2_{SM} + a_i \hat{v_i} + a_i {\mathcal M}_{ij} a_j.
\label{eqn:chi2}
\end{equation}

\vspace{0.15in} \noindent  We use the $\chi^2_{SM},\ \hat{v_i},$ and ${\mathcal M}_{ij}$ given by Refs. \cite{Han:2004az,Han:2005pr}, and $a_i$ are the operator coefficients, where the SM has been deformed: 

\begin{equation}
\mathcal{L}= \mathcal{L}_{SM}+a_i \mathcal{O}_i.
\end{equation}

\noindent The details of this method and a table of the relevant $a_i$ are given in the Appendix.  Some analyses of other little Higgs theories are given in refs. \cite{LHA}. 

By going to an extreme deconstruction of the 5d composite Higgs, we remove the $S$ parameter as a major constraint.  By imposing $\mathrm{SU(2)_C}$, we can suppress contributions to $T$ and take a limit where nonoblique corrections vanish.  The latter efficiently suppresses contributions to dimension 6 operators.  Thus, the theory is most constrained by oblique corrections.  After imposing EWSB and requiring the correct top mass, we can write the model in terms of five parameters:

\begin{equation*}
f, \theta', \theta_\lambda, \theta_y, \beta.
\end{equation*}

\noindent The vev of the nonlinear sigma model is set by $f$.  The $\theta$s are mixing angles for the gauge, plaquette, and top sector, respectively.  We define $\tan\beta$ as the ratio of the vev of the Higgs doublet which does not couple to SM fermions to the one that does.  For our analysis, we have set $\theta_\lambda$ to $\pi/4$, which corresponds to both plaquettes having the same coefficient.  This is justified by the approximate $\mathbb{Z}_4$ symmetry that rotates the links.  Small deviations from this value will give only suppressed contributions to $T$.  The authors of \cite{Chang:2003un} perform the full analytic analysis of PEW contributions.  We have independently checked those contributions except those resulting from the presence of two Higgs doublets, as these are standard \cite{Grant:1994ak, Bertolini:1985ia}.  

Despite having an explicit $\mathrm{SU(2)_C}$ in the theory, we still get contributions to $T$ from four different sectors: heavy gauge bosons, top sector, two Higgs doublets, and plaquette terms.  As mentioned, the last is subleading and we ignore it in the analysis.  The largest contribution comes from gauge bosons.  There is both a contribution from the incomplete cancellation of the $\mathrm{SU(2)_R}$ triplet, and a term from the custodial breaking Higgs vacuum:

\begin{equation}         
\Delta\rho_{\mathrm{gauge}}=-\frac{v^2}{64f^2}\sin^2 2\theta'+\frac{v^2}{64f^2}\sin^2 2\beta\sin^2\phi.
\end{equation}

\noindent For the original CMM, there are no one-loop contributions to $m_{xy}^2$, thus $\sin\phi=1$.  The largest contributions to the $S$ parameter come from having two Higgs doublets:

\begin{eqnarray}
\nonumber
\Delta S = \frac{1}{12 \pi}\Big[ 
\sin^2(\beta -\alpha) \log\frac{m^2_{H^0}}{m^2_{h^0}}
- \frac{11}{6} + \cos^2(\beta -\alpha) G(m^2_{H^0}, m^2_{A^0}, m^2_{H^\pm}) \\
+\sin^2(\beta -\alpha) G(m^2_{h^0}, m^2_{A^0}, m^2_{H^\pm})\Big], \vspace{0.2in}
\label{eqn:deltaS}
\end{eqnarray}

\noindent with

\begin{equation}
G(x,y,z)=\frac{x^2+y^2}{(x-y)^2}+\frac{(x-3y)x^2\log\frac{x}{z}-(y-3x)y^2\log\frac{y}{z}}{(x-y)^3}.
\end{equation}  

\noindent In the limit of large $f$, the first term in $S$ will dominate, and can even push $\Delta\chi^2$ past the 95\% C.L.  

Our remaining dimension six operators come from Higgs-Fermion and Fermion-Fermion currents.  We get the following operators and coefficients:

\begin{eqnarray}
\nonumber
O_{hf}^s = i(h^\dagger D^\mu h)(\overline{f}\overline{\sigma}^\mu f)+\mathrm{h.c.} & a_{hf}^s = \frac{Y \cos2\theta'\sin^2\theta'}{16f^2} \\
\nonumber
O_{hf}^t = i(h^\dagger \sigma^a D^\mu h)(\overline{f}\overline{\sigma}^\mu \sigma^a f)+\mathrm{h.c.} & a_{hf}^t = \frac{\cos2\theta\sin^2\theta}{32f^2} \\
\nonumber
O_{ff}^s = (\overline{f}_1\overline{\sigma}^\mu f_1)(\overline{f}_2\overline{\sigma}^\mu f_2)+\mathrm{h.c.} & a_{ff}^s = -\frac{Y_1 Y_2 \sin^4\theta'}{4f^2} \\
O_{ff}^t = (\overline{f}_1\overline{\sigma}^\mu \sigma^a f_1)(\overline{f}_2\overline{\sigma}^\mu \sigma^a f_2)+\mathrm{h.c.} & a_{ff}^t = -\frac{\sin^4\theta}{16f^2}.
\label{eqn:dim6}
\end{eqnarray}





\noindent Since all nonoblique contributions decouple in the limit that $\theta'\rightarrow 0$, the data generically prefer small $\theta'$.  However, we cannot take $\theta'$ too small without losing perturbativity at the SO(4) site.  Nonetheless, we still find acceptable areas of parameter space for $\theta'$ such that $g_4\lesssim3$.  There is also a perturbativity limit on $\theta_y$.  For $\theta_y\lesssim0.15$, the top quark parameter $y_2$ reaches strong coupling.  Once again, this does not significantly constrain the viable parameter region.  As mentioned above, for large enough $f$ the $\log\frac{m_{H^0}}{m_{h^0}}$ contribution to $S$ can become significant.  However, in general, new physics effects decouple as $f$ increases.  Furthermore, the new particles that cut off quadratic divergences get masses proportional to $f$.  For too large values, we can introduce a little hierarchy problem.  Thus, in general we are looking to take $f$ as small as possible.  Table \ref{tbl:fvalues} shows the 95\% CL on $f$ for a variety of $\theta',\ \theta_y,\ \mathrm{and}\ \beta$ values.  We have included upper limits on $f$ as well when these are $\mathcal{O}$(TeV).

\begin{table}[ht]
\begin{center}
\begin{tabular}{|c|c|c||c|}
\hline
$\theta'$ & $\beta$ & $\theta_y$ & 95\% CL on $f$ (GeV) \\\hline\hline
.12 & .5 & .2 & $>$465, $<$710 \\\hline
.12 & .3 & .15 & $>$380, $<$1260 \\\hline
.12 & .68 & .8 & $>$595, $<$880 \\\hline
.18 & .3 & .2 & 435 \\\hline
.18 & .5 & .4 & 725 \\\hline
.18 & .65 & .7 & $>$600, $<$875 \\\hline
.22 & .2 & .2 & 515 \\\hline
.22 & .54 & .4 & $>$590, $<$1395 \\\hline
.25 & .45 & .4 & 635 \\\hline
\end{tabular}
\end{center}
\caption{95\% CL on $f$.  Lower bounds except where indicated.}
\label{tbl:fvalues}
\end{table}

The original CMM paper \cite{Chang:2003un} did not perform a global analysis, and actually came up with overly conservative constraints.  The authors proposed a lower bound for $f$ at 700 GeV, and only in the limit that $g_4\rightarrow 3$.  In table \ref{tbl:fvalues} we see that we can get values of $f$ below 700 GeV (significantly in some cases), and even where $g_4\approx 1$.  The previous limit was determined by finding the $\theta',\ f$ that would satisfy limits for the most constraining operator.  By only comparing to an individual operator, one can get an artificially severe bound as one necessarily ignores cancellations between operators' contributions to observables.  For example, the 95\% CL on $S,\ T$ considered individually (11.8 TeV, 6.5 TeV) are nearly twice those arrived at when fit jointly (6.6 TeV, 3.7 TeV).  Thus, a weakening of constraints can occur from the addition of more operators.  

\begin{figure}[h]
\begin{center}
\includegraphics{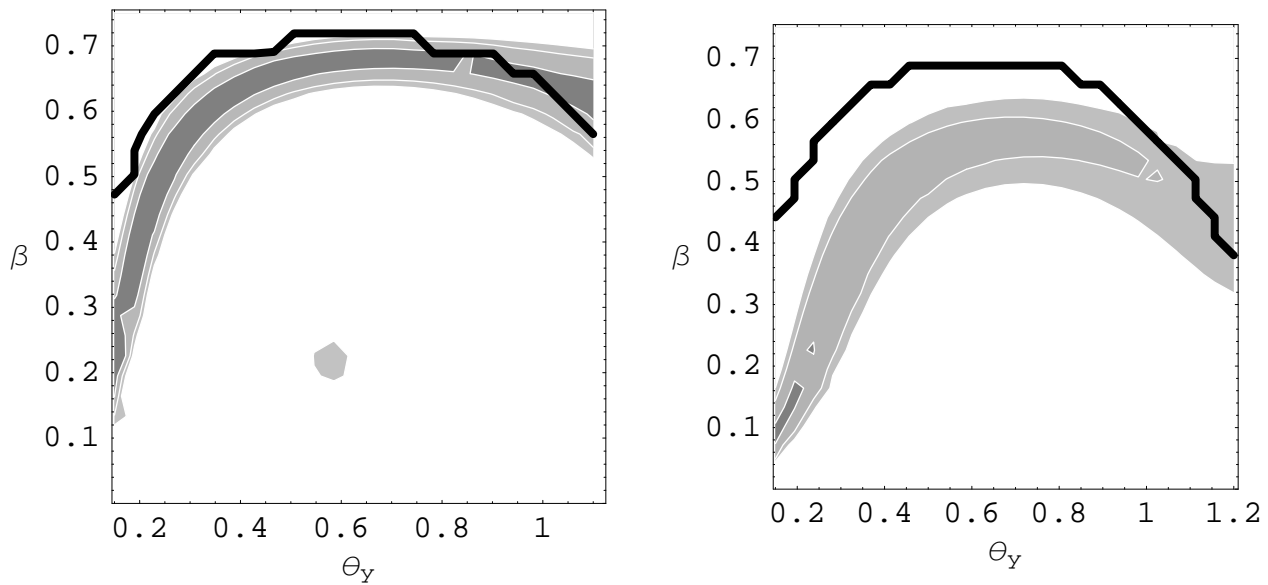}
\end{center}
\caption{$\Delta \chi^2$ contours as a function of $\beta$ vs. $\theta_y$ for $f=700$ GeV, $\theta'$=0.12 (left), and $\theta'$=0.25 (right).  The contours from darkest to lightest are at CL of 68\%, 95\% and 99\%. The regions above the thick black line are ruled out by a tachyonic physical Higgs.}
\label{fig:chi2thyb}
\end{figure}

\begin{figure}[h]
\begin{center}
\includegraphics{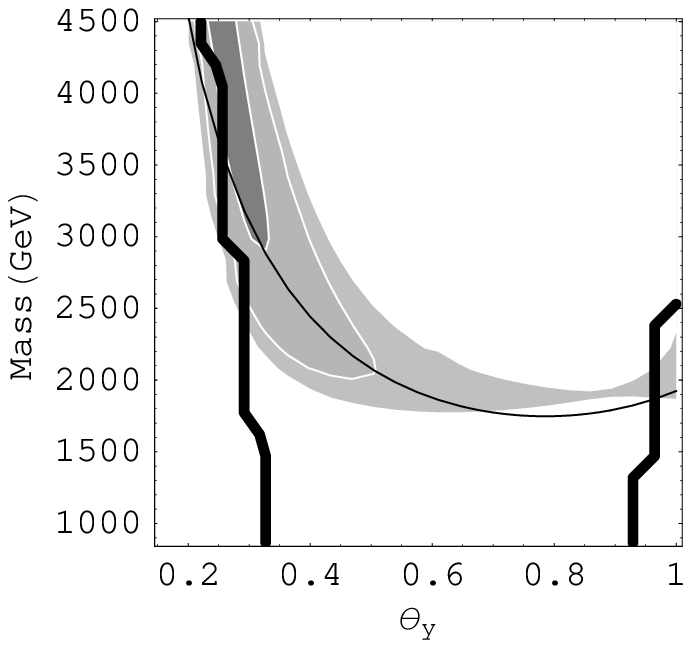}
\end{center}
\caption{$\Delta\chi^2$ contours as a function of $m_{B'}$ vs. $\theta_y$ for $f=600$ GeV, $\beta=0.28$.  The contours from darkest to lightest are at CL of 68\%, 95\% and 99\%.  The thin black curve is a plot of $m_{t'}$ vs. $\theta_y$ for the same values of $\beta$ and $f$.  The left and lower right regions enclosed by the thick black lines are ruled out by having a tachyonic physical Higgs mass.}
\label{fig:chi2thyMB}
\end{figure}

Figure \ref{fig:chi2thyb} shows the CL in $\beta-\theta_y$ space for two different values of $g_4$ at $f$=700 GeV.  \nolinebreak
\noindent We see that neither parameter is especially constrained even though we have imposed consistency with experiment, viable EWSB, correct value for $m_t$, and a non-tachyonic physical Higgs mass.  There are similarly sized acceptable regions for both $g_4$ near its perturbative limit ($g_4\approx 3$), and at $g_4\approx 1$.  However, in accordance with our intuition, the small $\theta'$ case has closer agreement to the data.  We can increase the size of the regions of acceptance by increasing $f$.  Figure \ref{fig:chi2thyMB}
shows the limits in terms of $m_{B'}$ and $m_{t'}$.  For small values of $\theta'$, we have that

\begin{equation}
m_{W'}=(1+2\theta'^2)m_{B'}.
\end{equation}  

\noindent Once again, we see no especially constrained parameters.  For our region of interest, the new heavy particles can take masses in the 2 to 5 TeV range.  We must have $m_{B'}\lesssim$4.5 TeV for perturbativity in $g_4$. For the values of $\beta$, $f$ plotted, the masses of $m_{B'}$ and $m_{t'}$ roughly track each other.  Even with new particles at the upper end of these mass scales, our Higgs potential is still natural, receiving contributions $\sim$200 GeV thanks to loop factor suppression.  

\section{Conclusion}
\label{sec:conc}

\hspace{0.15in} In studying theories of a PGB Higgs, we do well conceptually and numerically to go to four dimensions.  By deconstrucing the extra dimension and using holographic RG to run down, we can get an extreme deconstruction and describe the low energy physics in terms of a small moose.  Questions involving the symmetry of the theory, whether those protecting the Higgs mass or those in a particular breaking pattern become straightforward to answer.  In particular, we see that 5d warped models are just using collective symmetry breaking in the same way as the little Higgs to protect the Higgs potential.

In comparison with the data, the 4d model requires no tunings, nor was any individual parameter significantly constrained, nor did the data force us into a corner of parameter space where observables make opposite demands upon the theory.  The only additional symmetry needed for this was a gauged SO(4) to get an explicit custodial SU(2).  The fermion sector is modified by the presence of just one additional particle.  By contrast, the 5d warped analogue has all of the above difficulties, and requires the addition of a left-right parity and the placement of quarks into exotic representations of SO(4) to fix \zbb.   

As we search for the best realistic PGB Higgs model, little Higgs models with explicit $\mathrm{SU(2)_C}$ make a strong case (\cite{Chang:2003zn} offers a coset theory example).  We can turn any 5d warped model into a theory space little Higgs, instantly alleviating the $S$ problem and allowing for parametric separation in the Higgs potential between quartic and mass terms.  The CMM is the simplest possible way to include both $\mathrm{SU(2)_C}$ and an $\mathcal{O}(1)$ quartic without theory in theory space.  That being said, it is not without some aesthetic unpleasantness.  There are the $\epsilon$ and $\Omega$ plaquettes, as well as the spinor links to which fermions couple.  It would be interesting to see whether there is a more elegant model that performs as well in comparison to the data.  Given its success, we believe the CMM offers a standard against which to judge other theories of a PGB Higgs.           
\appendix

\section{$\chi^2$ Global Analysis}
\label{app:chi2}

\hspace{0.15in} The $\chi^2$ function for our analysis is parameterized as follows:

\begin{equation}
\chi^2=\chi^2_{SM} + a_i \hat{v_i} + a_i {\mathcal M}_{ij} a_j
\label{eqn:chi2a}
\end{equation}

\noindent In \cite{Han:2004az,Han:2005pr}, the authors obtain the $\chi^2_{SM},\ \hat{v_i},$ and ${\mathcal M}_{ij}$ from a global fit to the data of atomic parity violation, DIS, LEP, SLD, and Tevatron results.  The $a_i$ are a subset of the 80 $B$ and $L$ conserving dimension six operators of the SM \cite{Buchmuller:1985jz}.  One can eliminate several operators by demanding $CP$ conservation, and remove all those with only colored particles.  The latter step is justified by a lack of experimental constraints.  Lastly, we can impose various flavor symmetries.  We will use $\mathrm{U(3)^5}$, as our third generation makes the same contribution as the others at dimension six.  This leaves us with 20 operators for comparison to data.\footnote{Actually, there is a $\mathrm{21^{st}}$ operator, the triple gauge boson vertex correction, $(\mathcal{O}_W=\epsilon^{abc} W^{a\nu}_\mu W^{b\lambda}_\nu W^{c\mu}_\lambda)$.  However, since it is suppressed for the region of parameter space we are considering, we neglect it for the analysis.} 

We can express the 20 operators in terms of four model parameters: $f, \theta', \theta_y,$ and $\beta$.  The most complicated contributions are those from the Higgs sector to $S$ and $T$.  The Higgs potential is given in equation \ref{eqn:Higgspotl}.  In it, $m_x^2$ and $m_y^2$ are given by the Coleman-Weinberg potential.  Their leading contribution comes from the quadratic divergences cutoff by the heavy gauge, fermion, and scalar particles.  Thus, they are estimated as follows:

\begin{eqnarray}
\nonumber
m_y^2 &=& \frac{3}{64\pi^2}\left( 3g^2+g'^2 \right)\frac{4g'^2 f^2}{\sin^2\theta'}+\frac{6\,\lambda f^2}{\pi^2 \sin^2 2\theta_{\lambda}} \\
\ \\
m_x^2 &=& m_y^2-\frac{1}{32\pi^2}\frac{m_{\mathrm{top}}^4}{4f^2\, v^4 \cos^4 \beta} \sec^4 \theta_y \left( -8f^2 \csc^2 \theta_y + v^2 \cos^2 \beta \cos^2 \theta_y \sin\theta_y \right)^2.
\label{eqn:cwmasses}
\end{eqnarray}

\noindent  For our analysis, we set $\theta_{\lambda}=\frac{\pi}{4}$, as this is protected by the approximate $\mathbb{Z}_4$ link symmetry.  Following \cite{Chang:2003un}, minimization of the Higgs potential allows us to solve for EWSB parameters.

\begin{eqnarray}
2\lambda v^2 &=& (m_x^2 + m_y^2) \left( \frac{|b|}{m_x m_y }-1 \right)\\
\label{eqn:solveLambda}
\tan \beta &=& \frac{m_x}{m_y}\\
\tan 2\alpha &=& \left(1-\frac{2m_x m_y}{|b|}\right)\tan 2\beta.
\end{eqnarray}

\noindent Using equation \ref{eqn:solveLambda}, we can replace $\lambda$ in terms of our chosen four parameters.  

The Higgs sector contributions to $S$ and $T$ are functions of the physical masses squared, which are given by:      

\begin{eqnarray}
\nonumber
m^2_{A^0} &=& m_x^2 + m_y^2\\
\nonumber
m^2_{H^\pm} &=& m_x^2 + m_y^2 + 2 \lambda v^2 \\ 
\nonumber
m^2_{h^0} &=&  m^2_{H^\pm}\frac{\left(1-\sqrt{1-m^2_0/m^2_{H^\pm}} \right)}{2}\\
m^2_{H^0} &=&  m^2_{H^\pm}\frac{\left(1+\sqrt{1-m^2_0/m^2_{H^\pm}} \right)}{2},
\end{eqnarray}

\noindent with

\begin{eqnarray}
x = |b|/m_1 m_2
\hspace{0.5in} m_0^2 = \frac{8\lambda v^2 \sin^2 2\beta}{x}.
\end{eqnarray}

\noindent The Higgs contributions to $S$ (the only ones considered for our fit) are found in equation \ref{eqn:deltaS}.  They contribute as follows to $\delta\rho$:

\begin{eqnarray}
\nonumber
\delta \rho_{\mathrm{Higgs}} &=& \frac{\alpha}{16 \pi \sin^2\theta_{\text{w}} m^2_{W^\pm}}
\Big( F(m^2_{A^0},m^2_{H^\pm})\\
\nonumber
&&
\hspace{0.5in} 
+ \sin^2(\alpha -\beta) \big(F(m^2_{H^\pm}, m^2_{h^0}) -F(m^2_{A^0}, m^2_{h^0}) + 
\delta\hat{\rho}_{\text{SM}}(m^2_{H^0})\big)\\
&&
\hspace{0.5in} 
+ \cos^2(\alpha -\beta)
\big(F(m^2_{H^\pm}, m^2_{H^0}) -F(m^2_{A^0}, m^2_{H^0}) + \delta\hat{\rho}_{\text{SM}}(m^2_{h^0})\big)
\Big),
\end{eqnarray}

\noindent where

\begin{eqnarray}
F(x,y) &=& \half(x + y) - \frac{xy}{x -y} \log\frac{x}{y}\\
\nonumber
\delta \hat{\rho}_{\text{SM}}(m^2)&=&  F(m^2,m^2_{W^\pm}) - F(m^2,m^2_{Z^0})\\
&&+ \frac{4 m^2 m^2_{W^\pm}}{m^2 - m^2_{W^\pm}} \log \frac{m^2}{m^2_{W^\pm}}
- \frac{4 m^2 m^2_{Z^0}}{m^2 - m^2_{Z^0}} \log \frac{m^2}{m^2_{Z^0}}.
\end{eqnarray}

\noindent There is an additional one-loop contribution to $\delta\rho$ from the top sector (for our analysis we use $m_{\mathrm{top}}$ = 173 GeV):

\begin{eqnarray}
\delta \rho_{\mathrm{top}} &\simeq& 
\frac{3\, m_{\mathrm{top}}^2\, \cos^2 \beta \sin^4 \theta_y }{64 \pi^2 f^2}
\Big(
\tan^2 \theta_y  -2 
\big( \log\frac{v^2 \sin^2 \theta_y \cos^2 \theta_y \cos^2\beta}{4 f^2}
+ 1\big) \Big).
\end{eqnarray}

Combining these with the Higgs-fermion and fermion-fermion coefficients of equation \ref{eqn:dim6}, we can write down the 20 dimension six operator coefficients in terms of $f, \theta', \theta_y,$ and $\beta$.  We have replaced $\theta$ in terms of $\theta'$, using $\sin\theta \approx \sqrt{3}\sin\theta'$.

\begin{equation}
\begin{array}{cclccl}
a_h&=&-\frac{2}{v^2}\left(\delta\rho_{\mathrm{gauge}}+\delta\rho_{\mathrm{Higgs}}+\delta\rho_{\mathrm{top}} \right)+\delta\hat{a}_h, \\ 
\\
a_{wb}&=&\frac{1}{2\sin 2\theta_w}\frac{\alpha}{v^2}\Delta S + \delta\hat{a}_{wb}, \\
\\
a^t_{hl}&=&\frac{3\sin^2\theta' \left(3\cos 2\theta' - 2 \right)}{32f^2} ,
&a^t_{hq}&=&
\frac{3\sin^2\theta' \left(3\cos 2\theta' - 2 \right)}{32f^2}, \\
a^s_{hl}&=&-\frac{\sin^2\theta' \cos 2\theta'}{32f^2},
& a^s_{hq}&=& \frac{\sin^2\theta' \cos 2\theta'}{96 f^2}, \\
a_{hu}&=& \frac{\sin^2\theta' \cos 2\theta'}{24f^2},
& a_{hd}&=&-\frac{\sin^2\theta' \cos 2\theta'}{48f^2}, \\
a_{he}&=& -\frac{\sin^2\theta'\cos 2\theta'}{16f^2},\\
\\
a^t_{ll}&=&-\frac{9 \sin^4 \theta'}{8f^2}, & a^t_{lq}&=&-\frac{9 \sin^4 \theta'}{16f^2},\\
a^s_{ll}&=& -\frac{\sin^4 \theta'}{8f^2},
& a^s_{lq}&=&\frac{\sin^4 \theta'}{48f^2},\\
a_{le}&=&-\frac{\sin^4 \theta'}{8f^2},
& a_{qe}&=&\frac{\sin^4 \theta'}{24f^2},\\
a_{lu}&=&\frac{\sin^4 \theta'}{12f^2},
& a_{ld}&=&-\frac{\sin^4 \theta'}{24f^2},\\
a_{ee}&=&-\frac{\sin^4 \theta'}{2f^2},
& a_{eu}&=&\frac{\sin^4 \theta'}{6f^2},\\
a_{ed}&=&-\frac{\sin^4 \theta'}{12f^2}
\end{array}
\end{equation}

\vspace{0.1in}
\noindent The terms $\delta\hat{a}_h$ and $\delta\hat{a}_{wb}$ are the logarithmic contributions that come from having a Higgs mass shifted from the global minimum of the precision data:

\begin{equation}
\delta\hat{a}_h \approx \frac{3\alpha}{8 \pi\,v^2 \cos^2 \theta_w}\log\left(\frac{m_h^2}{113\;\mathrm{GeV}^2}\right), \quad \delta\hat{a}_{wb} \approx \frac{\alpha}{48 \pi\,v^2 \cos \theta_w \sin \theta_w}\log\left(\frac{m_h^2}{113\; \mathrm{GeV}^2}\right).
\end{equation}

\noindent For much of parameter space, they give the dominant contributions to $\chi^2$.

\section*{Acknowledgments}
Thank you to Nima Arkani-Hamed for inspiring and advising the project.  Thanks to Roberto Contino for many useful insights into his papers, Spencer Chang and Jay Wacker for their feedback, and to Clifford Cheung for illuminating discussions on little Higgs theories and deconstruction.  

\begingroup\raggedright

\endgroup

\end{document}